\documentclass[seceq]{ptptex}
\usepackage{wrapft}
\usepackage{graphicx}

\newcommand{\bra}[1]{\langle {#1} |}     %%
\newcommand{\ket}[1]{| {#1} \rangle}     %%
\newcommand{\kket}[1]{| {#1} \rangle\!\rangle}     %%
\newcommand{\rrket}[1]{| {#1} ))}     %%
\newcommand{\dket}[1]{|\!| {#1} \rangle}     %%
\newcommand{\ma}[1]{\stackrel{\tiny\circ} {#1}} %%
\newcommand{\maru}[1]{\breve{#1}} %%

\def\beq{\begin{eqnarray}}
\def\eeq{\end{eqnarray}}
\def\bsub{\begin{subequations}}
\def\esub{\end{subequations}}
\def\b{\begin{equation}}
\def\bs{\begin{split}}
\def\es{\end{split}}
\def\e{\end{equation}}
\begin{document}

\title{Beyond the Schwinger boson representation of the $su(2)$-algebra. II
}
\subtitle{Some theoretical features of new boson representation and connections to 
the other boson representations
}

\author{%       %Use \sc for the family name
Yasuhiko {\sc Tsue},$^{1}$, Constan\c{c}a {\sc Provid\^encia},$^{2}$
Jo\~ao da {\sc Provid\^encia},$^{2}$  and Masatoshi {\sc Yamamura}$^{3}$
}

\inst{%         %Affiliation, neglected when [addenda] or [errata]
$^{1}$Physics Division, Faculty of Science, Kochi University, Kochi 780-8520, Japan\\
$^{2}$Departamento de F\'{i}sica, Universidade de Coimbra, 3004-516 Coimbra, 
Portugal\\
$^{3}$Department of Pure and Applied Physics, 
Faculty of Engineering Science, Kansai University, Suita 564-8680, Japan
}

\abst{
Concerning the new boson representation presented in Part I, it is 
proved that this representation obeys the $su(2)$-algebra in a certain subspace 
in the whole boson space constructed by the Schwinger boson representation of 
the $su(1,1)$-algebra. 
Some other problems related to this representation are discussed. 
}

%\subjectindex{xxxx, xxx}

\maketitle

\section{Introduction}

The present paper, Part II, is the continuation of the previous work referred to 
as (I) \cite{1}. 
In (I), we presented a new boson representation of the $su(2)$-algebra. 
In the same scheme as that in the Schwinger boson representation \cite{2}, 
three generators in our case are also expressed in terms of two kinds of bosons. 
Concrete forms can be seen in the relations (I.3.7) and (I.3.8c) or 
(I.3.9) and (I.3.10). 
The operator for the magnitude of the $su(2)$-spin is given in the relation (I.3.10): 
${\hat {\cal S}}=({\hat a}^*{\hat a}-{\hat b}^*{\hat b})/2+C_m$. 
Here, $C_m$ denotes a certain constant which is 
appropriately chosen. 
On the other hand, in the Schwinger representation, ${\hat S}$ is given as 
${\hat S}=({\hat a}^*{\hat a}+{\hat b}^*{\hat b})/2$, 
which is seen in the relation (I.3.2b). 
It is an essential difference between the Schwinger and our representation. 
In (I), we promised to prove that our new boson representation obeys the 
$su(2)$-algebra in Part II, i.e., 
the present paper. 
It is our central task of this paper. 
For this proof, we must consider the subspace (I.2.8) in the whole space given in 
the Schwinger boson representation of the $su(1,1)$-algebra. 
In this subspace, our representation satisfies the $su(2)$-algebra. 
Of course, we discuss the algebras in the spaces which are orthogonal to the 
subspace (I.2.8).

We know two forms of the boson representations of the $su(2)$-algebra. 
One is, of course, the Schwinger representation and the other is the 
Holstein-Primakoff representation \cite{3}. 
It may be interesting to investigate the connection of ours to the other two. 
As will be shown in \S 4, the Holstein-Primakoff representation can be derived 
rather easily from ours. 
However, the relation between the Schwinger representation and ours is rather complicated. 
In common with the two, they are formulated in terms of two kinds of bosons. 
But, ours contains one parameter $C_m$, which can be seen in the relations (I.3.9) 
and (I.3.10). 
For the understanding of $C_m$, the pairing model in many-fermion system 
is an instructive example. 
In this model, we have $2C_m=4\Omega_0$, the total number of the single-particle states, 
which is shown in p.11 of (I). 
The above example suggests us that $C_m$ is regarded as the parameter 
expressing the size of the system under consideration. 
In \S 5, we can show that, as a natural consequence, the magnitude of the 
$su(2)$-spin $s$ can change in the range $s=0,\ 1/2,\cdots ,\ C_m/2-1,\ C_m/2$, 
which is consistent to the well-known formula in the pairing model. 
In the Schwinger representation, there does not exist such restriction and, then, 
$s=0,\ 1/2, \cdots,\ \infty$. 
Certainly, if $C_m\rightarrow \infty$, we can show that ours is reduced to 
the Schwinger representation. 
In \S 5, the above will be discussed. 
In the pseudo $su(1,1)$-algebra, we introduced $t_m$ as the maximum value of 
$t_0$ for a given $t$ and, as a possible example, we gave the form 
$t_m=C_m+1-t$ (see (I.2.16a)). 
However, this form is not unique and there exist infinite possibilities 
and in \S 6, we will give another example; $t_m=3t-1$.

Next section is central part of this paper. 
In this section, it is proved that, in the subspace (I.2.8), ${\hat {\cal S}}_{\pm,0}$ introduced 
in the relation (I.3.9) obey the $su(2)$-algebra and ${\hat {\cal S}}$ given in the 
relation (I.3.10) plays a role of the magnitude of the $su(2)$-spin. 
In \S 3, raising and lowering operator for the magnitude of the $su(2)$-spin 
are discussed, respectively. 
Finally, in \S 6, as the concluding remarks, two problems are treated. 
One is related to the algebras in the space orthogonal to the subspace (I.2.8). 
Partly, this problem is discussed in \S 2. 
The other is concerned with another example of $t_m$; $t_m=3t-1$.

\setcounter{equation}{0}

\section{The boson representation of the $su(2)$-algebra presented in Part I}

\begin{figure}[t]
\begin{center}
\includegraphics[height=10.5cm]{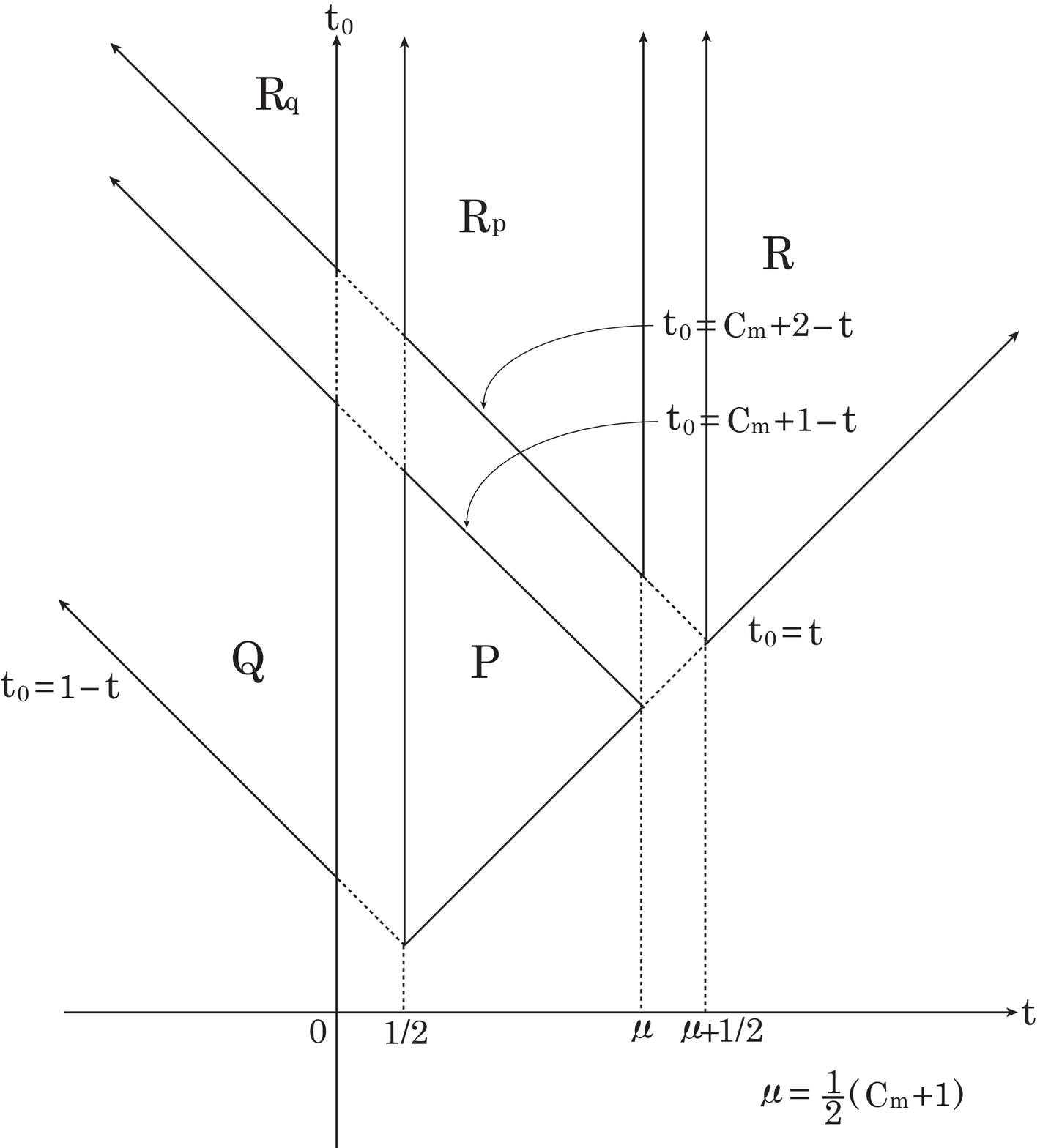}
\caption{The subspaces $P$, $Q$, $R_p$, $R_q$ and $R$ are shown on the $t$-$t_0$ plane.
}
\label{fig:subspace}
\end{center}
\end{figure}

In \S I-3, we formulated a new boson representation of the $su(2)$-algebra. 
After giving the relation (I.3.8c) and (I.3.22), we mentioned that 
our representation holds in the space (I.2.8) as a subspace of the whole space (I.2.5). 
But, this mentioning was presented without any explanation. 
Main aim of this section is to formulate this mentioning strictly. 
We sketch out the space (I.2.8) and other subspace on the $t$-$t_0$ plane. 
As was already mentioned, we have been interested in the space obeying the condition (I.2.8), 
which is a subspace of the whole space specified by the condition (I.2.5). 
Figure \ref{fig:subspace} shows various subspaces for the case ${\hat T}_m=C_m+1-{\hat T}$. 
This formula was given in the range $1/2 \leq t \leq \mu\ (=(C_m+1)/2)$ with a certain reason discussed in \S I-2. 
We assume that the above formula for ${\hat T}_m$ is also useful in the range 
$-\infty < t \leq 0$ and, in this paper, the case ${\hat T}_m=C_m+1-{\hat T}$ is treated exclusively. 
In our present scheme, the whole space is divided into five subspaces $P$, $Q$, $R_p$, $R_q$ and $R$, which 
are shown in Fig.\ref{fig:subspace}. 
Of course, we are mainly interested in $P$. 
With the above-mentioned point in mind, 
the $su(2)$-algebra was formulated in \S I-3. 
The expression of ${\hat {\cal S}}_{\pm,0}$ and ${\hat {\cal S}}$ are 
given in the relations (I.3.7) and (I.3.8) or (I.3.8c) and (I.3.10). 
In this section, we will examine some properties of these expressions in $P$ and $Q$.

%%%%%%%%%%%%%%%%%%%%%%%%%%%%%%%%%%%%%%%%%%%%%%%%%%%%%%%%%%%%%%%%%%%%%%
%%%%%%%%%%%%%%%%%%%%%%%%%%%%%%%%%%%%%%%%%%%%%%%%%%%%%%%%%%%%%%%%%%%%%%%%

For the preparation of the main discussion, first, we make a list of the relations. 
The relations (I.3.7a) and (I.3.8c) are rewritten as 
\beq
& &{\hat T}=\frac{1}{2}\left(C_m+1\right)-{\hat {\cal S}}\ , \qquad
{\hat T}_0=\frac{1}{2}\left(C_m+1\right)+{\hat {\cal S}}_0\ , 
\label{a1}\\
{\rm i.e.,}\quad 
& &t=\frac{1}{2}\left(C_m+1\right)-s\ , \qquad
t_0=\frac{1}{2}\left(C_m+1\right)+s_0\ . 
\label{a2}
\eeq
The relations (I.3.9a) and (I.3.9b) are rewritten as 
\bsub\label{a3}
\beq
& &{\hat {\cal S}}_+={\hat a}^*{\hat b}^*\cdot\left(\sqrt{C_m+1-{\hat {\cal S}}+{\hat {\cal S}}_0+\epsilon}\right)^{-1}\cdot
\sqrt{{\hat {\cal S}}-{\hat {\cal S}}_0}\ , 
\label{a3a}\\
& &{\hat {\cal S}}_-=\sqrt{{\hat {\cal S}}-{\hat {\cal S}}_0}\cdot\left(\sqrt{C_m+1-{\hat {\cal S}}+{\hat {\cal S}}_0+\epsilon}\right)^{-1}\cdot
{\hat b}{\hat a}\ . 
\label{a3b}
\eeq
\esub
The minimum weight state denoted by $\ket{s,\sigma}$ obeys the conditions
\beq
& &{\hat {\cal S}}_-\ket{s,\sigma}=0\ , \qquad
{\rm i.e.,}\qquad
{\hat b}{\hat a}\ket{s,\sigma}=0\ , 
\label{a4}\\
& &{\hat {\cal S}}\ket{s,\sigma}=s\ket{s,\sigma}\ , \qquad
{\hat {\cal S}}_0\ket{s,\sigma}=\sigma\ket{s,\sigma}\ . 
\label{a5}
\eeq
The relations (\ref{a4}) and (\ref{a5}) with (\ref{a3b}) give us 
\beq
& & P\ ; \ \ \ket{s,\sigma}=({\hat b}^*)^{C_m-2s}\ket{0}\ , \qquad \sigma=-s\ , 
\label{a6}\\
& & Q\ ; \ \ \ket{s,\sigma}=({\hat a}^*)^{2s-C_m}\ket{0}\ , \qquad \sigma=s-C_m\ . 
\label{a7}
\eeq
Here, the normalization constants are omitted. 
Since $C_m-2s\geq 0$ and $2s-C_m \geq 1$, we have the inequalities 
\beq
& & P\ ;\quad s\leq \frac{1}{2}C_m\ , 
\label{a8}\\
& & Q\ ; \quad s\geq \frac{1}{2}\left(C_m+1\right)\ . 
\label{a9}
\eeq
By operating ${\hat {\cal S}}_+$ successively on the state $\ket{s,\sigma}$, we are able to 
obtain the eigenstate of ${\hat {\cal S}}$ and ${\hat {\cal S}}_0$ with the eigenvalues $s$ and $s_0$, 
respectively, in the form
\beq
& &P\ ; \ \ \ket{s,s_0}=\left({\hat {\cal S}}_+\right)^{s+s_0}\ket{s, -s}\ , 
\label{a10}\\
& &Q\ ; \ \ \ket{s,s_0}=\left({\hat {\cal S}}_+\right)^{C_m-s+s_0}\ket{s,s-C_m}\ . 
\label{a11}
\eeq
Here, the normalization constants are omitted. 
Since $s+s_0 \geq 0$ and $C_m-s+s_0 \geq 0$, we have 
\beq
& &P\ ; \ \ s_0 \geq -s\ , 
\label{a12}\\
& &Q\ ; \ \ s_0\geq s-C_m\ . 
\label{a13}
\eeq
Our system has the maximum weight state denoted as $\ket{s,\sigma_m}$.
 It satisfies 
\beq
& &{\hat {\cal S}}_+\ket{s,\sigma_m}=0\ , \qquad {\rm i.e.,}\qquad
\sqrt{{\hat {\cal S}}-{\hat {\cal S}}_0}\ \ket{s,\sigma_m}=0\ , 
\label{a14}\\
& &{\hat {\cal S}}\ket{s,\sigma_m}=s\ket{s,\sigma_m}\ , \qquad
{\hat {\cal S}}_0\ket{s,\sigma_m}=\sigma_m\ket{s,\sigma_m}\ . 
\label{a15}
\eeq
The relation (\ref{a14}) and (\ref{a15}) give us
\beq\label{a16}
P\ \ {\rm and}\ \ Q\ ; \quad \sigma_m=s\ . 
\eeq
Then, the quantum number $s_0$ satisfies 
\beq
& &P\ ; \ \ -s\leq s_0 \leq s\ , \qquad {\rm i.e.,}\qquad s_0=-s,\ -s+1,\cdots ,\ s-1,\ s, 
\label{a17}\\
& &Q\ ; \ \ s-C_m \leq s_0 \leq s\ , \qquad {\rm i.e.,}\qquad s_0=s-C_m,\ s-C_m+1,\cdots ,\ s-1,\ s. \qquad
\label{a18}
\eeq
We can see that ${\hat {\cal S}}_{\pm,0}$ obey the $su(2)$-algebra in $P$, but do not in $Q$. 
The relations (\ref{a8}) and (\ref{a9}) teach us that the upper and the lower limit of $s$ in $P$ and in $Q$, respectively, exist.
Therefore, it may be interesting to show the lower and the upper limit of $s$ in $P$ and in $Q$, respectively. 
For this task, the first of the relation (\ref{a2}) is useful. 
As can be seen in Fig.\ref{fig:subspace}, the quantum number $t$ in $P$ and in $Q$ obeys 
\beq
& &P\ ; \ \ t=1/2,\ 1,\ 3/2,\cdots ,\ \mu-1/2,\ \mu,
\label{a19}\\
& &Q\ ; \ \ t=0, -1/2,\ -1,\cdots ,\ -\infty.
\label{a20}
\eeq
Combining the relations (\ref{a19}) and (\ref{a20}) with the relation (\ref{a2}), we have the following: 
\beq
& &P\ ; \ \ s=\frac{1}{2}C_m,\ \frac{1}{2}(C_m-1), \ \frac{1}{2}(C_m-2),\cdots ,\ \frac{1}{2},\ 0, 
\label{a21}\\
& &Q\ ; \ \ s=\frac{1}{2}(C_m+1), \ \frac{1}{2}(C_m+2),\ \frac{1}{2}(C_m+3),\cdots ,\ \infty.
\label{a22}
\eeq
In $P$, there exists the lower limit $s=0$, but in $Q$, the upper limit is $\infty$.

Next, we investigate the commutation relations for ${\hat {\cal S}}_{\pm,0}$ and ${\hat {\cal S}}$. 
For the above discussion, the following played a central role: 
\beq\label{a23}
[\ {\hat {\cal S}}_0\ , \ {\hat {\cal S}}_{\pm}\ ]=\pm{\hat {\cal S}}_{\pm}\ , \qquad
[\ {\hat {\cal S}}\ , \ {\hat {\cal S}}_{\pm,0}\ ]=0 \ .
\eeq 
Of course, the above relation is useful both in $P$ and in $Q$. 
Our problem is to investigate the relation $[\ {\hat {\cal S}}_+\ , \ {\hat {\cal S}}_-\ ]$. 
Direct calculation gives us the form
\beq
& &[\ {\hat {\cal S}}_+\ , \ {\hat {\cal S}}_-\ ]=2\left({\hat {\cal S}}_0+\Delta{\hat {\cal S}}_0\right)\ , 
\label{a24}\\
& &\Delta{\hat {\cal S}}_0=\Delta{\hat {\cal S}}_0^{(+)}-\Delta{\hat {\cal S}}_0^{(-)}\ , \nonumber\\
& &\Delta{\hat {\cal S}}_0^{(+)}=\frac{\epsilon}{2}\frac{\left({\hat T}_0-{\hat T}+1\right)\left(C_m+1-{\hat T}_0-{\hat T}\right)}{{\hat T}_0+{\hat T}+\epsilon}\ , \nonumber\\
& &
\Delta{\hat {\cal S}}_0^{(-)}=\frac{\epsilon}{2}\frac{\left({\hat T}_0-{\hat T}\right)\left(C_m+2-{\hat T}_0-{\hat T}\right)}{{\hat T}_0+{\hat T}-1+\epsilon}\ . 
\label{a25}
\eeq
Apparently, $[\ {\hat {\cal S}}_+\ , \ {\hat {\cal S}}_-\ ]$ does not satisfy the commutation relation in the $su(2)$-algebra. 
As can be seen in Fig.\ref{fig:subspace}, $t_0$ and $t$ obey the inequality $t_0+t>0$ both in $P$ and in $Q$. 
It indicates that the term $({\hat T}_0+{\hat T})$ appearing in the denominator of $\Delta{\hat {\cal S}}_0^{(+)}$ is positive-definite and, then, we have 
\beq\label{a26}
P\ {\rm and}\ Q\ ; \ \ \Delta{\hat {\cal S}}_0^{(+)}\longrightarrow 0\ , \quad (\epsilon \rightarrow 0)\ . 
\eeq
In the case $\Delta{\hat {\cal S}}_0^{(-)}$, it may be necessary to investigate  if the operator $({\hat T}_0+{\hat T}-1)$ appearing in the denominator is positive-definite or not. 
For this aim, the condition $t_0+t-1=0$ should 
be examined: 
In $P$, $t_0=t=1/2$ and in $Q$, $t_0=-t+1$ $(t\leq 0)$. 
Figure \ref{fig:subspace} gives us these relations. 
We notice the term $({\hat T}_0-{\hat T})$ appearing in the numerator of $\Delta{\hat {\cal S}}_0^{(-)}$. 
In this case, $t_0-t=1/2-1/2=0$. 
Therefore, the following result is derived: 
\beq\label{a27}
P\ ; \ \ \Delta{\hat {\cal S}}_0^{(-)}\longrightarrow 0\ , \quad (\epsilon \rightarrow 0)\ . 
\eeq
On the other hand, there does not exist the term which leads to $\Delta{\hat {\cal S}}_0^{(-)}\rightarrow 0$, $(\epsilon\rightarrow 0)$. 
Therefore, we have 
\beq
Q\ ; \ \ & &\Delta{\hat {\cal S}}_0^{(-)}\longrightarrow \frac{1}{2}\left({\hat T}_0-{\hat T}\right)\left(C_m+2-{\hat T}_0-{\hat T}\right)\cdot {\hat Q}_0\ , 
\quad (\epsilon\rightarrow 0)\ , 
\label{a28}\\
& &{\hat Q}_0=\sum_{t\leq 0}\ket{t,t_0=-t+1}\bra{t,t_0=-t+1}\nonumber\\
& &\quad\ \ 
=\sum_{s\geq 1/2\cdot(C_m+1)}\ket{s,\sigma}\bra{s,\sigma}\ . \quad (\sigma=s-C_m) 
\label{a29}
\eeq
For the operator ${\hat {\mib {\cal S}}}^2$, we have 
\beq\label{a30}
{\hat {\mib {\cal S}}}^2={\hat {\cal S}}\left({\hat {\cal S}}+1\right)-\frac{1}{2}\left(\Delta{\hat {\cal S}}_0^{(+)}+\Delta{\hat {\cal S}}_0^{(-)}\right)\ . 
\eeq
Therefore, for $\epsilon \rightarrow 0$, ${\hat {\mib {\cal S}}}^2$ is expressed in the form 
\beq
& &P\ ; \ \ {\hat {\mib {\cal S}}}^2\longrightarrow {\hat {\cal S}}\left({\hat {\cal S}}+1\right)\ ,  
\label{a31}\\
& &Q\ ; \ \ {\hat {\mib {\cal S}}}^2\longrightarrow {\hat {\cal S}}\left({\hat {\cal S}}+1\right)-\frac{1}{2}\left({\hat T}_0-{\hat T}\right)
\left(C_m+2-{\hat T}_0-{\hat T}\right)\cdot {\hat Q}_0\ . 
\label{a32}
\eeq
The above consideration teaches us that ${\hat {\cal S}}_{\pm,0}$ obey the $su(2)$-algebra in $P$, but 
they does not obey the $su(2)$-algebra in $Q$. 
It may be permitted to call the algebra in $Q$ the pseudo $su(2)$-algebra. 
With the use of the commutation relation (\ref{a24}), we can determine the normalization constants of the states (\ref{a6}) and (\ref{a7}). 
For this aim, the following formula is useful: 
\beq\label{a33}
\bra{s,\sigma}\left({\hat {\cal S}}_-\right)^n\cdot \left({\hat {\cal S}}_+\right)^n\ket{s,\sigma}
=(-)^n n!\frac{\Gamma(2\sigma+n)}{\Gamma(2\sigma)}\bra{s,\sigma}s,\sigma\rangle\ . 
\eeq
Combining the formula (\ref{a33}) with the relations (\ref{a6}), (\ref{a7}), (\ref{a10}) and (\ref{a11}), we can determine 
the norms of the states (\ref{a10}) and (\ref{a11}) in the following form :
\beq
& &P\ ; \ \ \bra{s,s_0}s,s_0\rangle=\frac{(2s)!(s+s_0)!}{(s-s_0)!}\cdot (C_m-2s)!\ , 
\label{a34}\\
& &Q\ ; \ \ \bra{s,s_0}s,s_0\rangle=(-)^{C_m-s+s_0}\frac{(C_m-s+s_0)!\Gamma(s+s_0-C_m)}{\Gamma(2(s-C_m))}\cdot(2s-C_m)!\ . 
\label{a35}
\eeq
With the aid of the relations (\ref{a34}) and (\ref{a35}), we are able to obtain the normalized $\ket{s,s_0}$. 
Clearly, the relation (\ref{a34}) is the same as that in the $su(2)$-algebra. 
Of course, $(C_m-2s)!$ is derived under the minimum weight state (\ref{a6}). 
The above is the supplementary explanation of our boson representation of the $su(2)$-algebra. 
Certainly, in $P$, our representation obeys the $su(2)$-algebra. 
Concerning the subspaces $R_p$, $R_q$ and $R$, we will discuss in \S 6.

\setcounter{equation}{0}

\section{Raising and lowering operator for the magnitude of the $su(2)$-spin}

In this section, we will discuss a role of the original Schwinger representation (I.3.1) in our present one. 
It is shown that the operators ${\hat S}_+$ and ${\hat S}_-$ play the role of the raising and the lowering 
operator for the magnitude of the $su(2)$-spin, respectively. 
First, we notice that the relation between $(t,t_0)$ and $(s,s_0)$ is given as 
\beq\label{a48}
t=\frac{1}{2}(C_m+1)-s\ , \qquad
t_0=\frac{1}{2}(C_m+1)+s_0\ . 
\eeq
The form (\ref{a48}) is derived from the relations (I.3.7a) and (I.3.8c) with 
the relation (I.2.16a). 
The space is characterized by the conditions $t \leq t_0 \leq C_m+1-t$ and 
$1/2 \leq t \leq \mu$, which leads us to the following conditions 
for $(s,s_0)$: 
\bsub\label{a49}
\beq
& &-s \leq s_0 \leq s\ , 
\label{a49a}\\
& &0 \leq s \leq \frac{1}{2}C_m\ . 
\label{a49b}
\eeq
\esub
%%%%%%%%%%%%%%%%%%%%%%%%%%%%%%%%%%%%%%%%%%%%%%%%%%%%%%%%%%%%%%%%%%%%%%
\begin{figure}[t]
\begin{center}
\includegraphics[height=6.5cm]{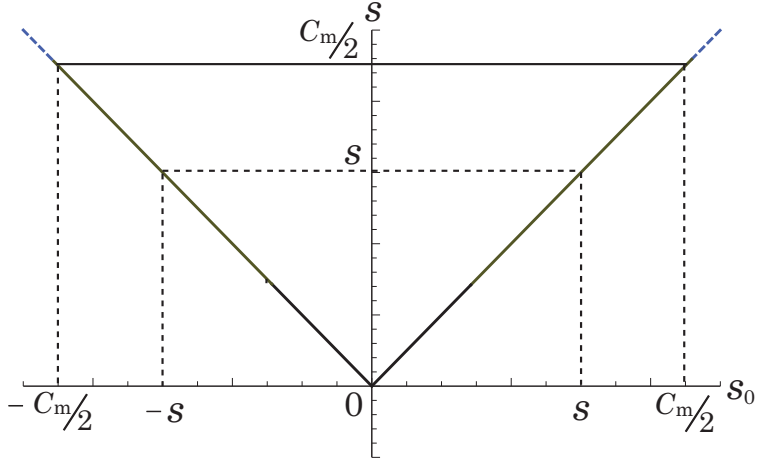}
\caption{The condition (\ref{a49}) is depicted on the $s$-$s_0$ plane.
}
\label{fig:ss0}
\end{center}
\end{figure}
%%%%%%%%%%%%%%%%%%%%%%%%%%%%%%%%%%%%%%%%%%%%%%%%%%%%%%%%%%%%%%%%%%%%%%%%
The condition (\ref{a49}) can be shown in Fig.\ref{fig:ss0}. 
With the use of the relation (\ref{a48}), the state $\ket{t,t_0}$ in $P$ can be 
expressed in the form 
\beq\label{a50}
& &\ket{t,t_0}=\frac{1}{\sqrt{(t_0-t)!(t_0+t-1)!}}({\hat a}^*)^{t_0-t}({\hat b}^*)^{t_0+t-1}\ket{0}\nonumber\\
&=&\ket{s,s_0}=\frac{1}{\sqrt{(s+s_0)!(C_m-s+s_0)!}}({\hat a}^*)^{s+s_0}({\hat b}^*)^{C_m-s+s_0}\ket{0}\ . 
\eeq
The state $\dket{s,s_0}$ in the Schwinger representation is given as 
\beq\label{a51}
\dket{s,s_0}=\frac{1}{\sqrt{(s+s_0)!(s-s_0)!}}({\hat a}^*)^{s+s_0}({\hat b}^*)^{s-s_0}\ket{0}\ . 
\eeq

The state (\ref{a50}) gives us the relation 
\bsub\label{a52}
\beq
& &{\hat S}_+\ket{s,s_0}={\hat a}^*{\hat b}\ket{s,s_0}
=\sqrt{(s+s_0+1)(C_m-s+s_0)}\ket{s+1,s_0}\ , 
\label{a52a}\\
& &{\hat S}_-\ket{s,s_0}={\hat b}^*{\hat a}\ket{s,s_0}
=\sqrt{(s+s_0)(C_m-s+s_0+1)}\ket{s-1,s_0}\ , 
\label{a52b}\\
& &{\hat S}_0\ket{s,s_0}=\frac{1}{2}({\hat a}^*{\hat a}-{\hat b}^*{\hat b})\ket{s,s_0}
=\left(s-\frac{1}{2}C_m\right)\ket{s,s_0}\ . 
\label{a52c}
\eeq
\esub
Here, ${\hat S}_{\pm,0}$ denote the generators of the $su(2)$-algebra in the 
Schwinger representation. 
In the relations (\ref{a52a}) and (\ref{a52b}), we can see that 
${\hat S}_+$ and ${\hat S}_-$ play the role of the raising and the lowering operator, 
respectively, not for $s_0$ but for $s$. 
However, we must notice the case $s=C_m/2$. 
Operation of ${\hat S}_+$ on the state $\ket{C_m/2, s_0}$ should vanish, because $C_m/2+1$ 
does not belong to the space $P$. 
In order to overcome this trouble, 
we define the following operators: 
\bsub\label{a53}
\beq
& &{\hat \Sigma}_+={\hat S}_+\!\cdot\!\sqrt{\frac{1}{2}C_m-{\hat {\cal S}}}\!\cdot\!
\left(\sqrt{\frac{1}{2}C_m-{\hat {\cal S}}+\epsilon}\right)^{-1}\ , \nonumber\\
& &
\label{a53a}\\
& &{\hat \Sigma}_-=
\left(\sqrt{\frac{1}{2}C_m-{\hat {\cal S}}+\epsilon}\right)^{-1}\!\!
\cdot\!
\sqrt{\frac{1}{2}C_m-{\hat {\cal S}}}\cdot{\hat S}_- , \nonumber\\
& &
\label{a53b}\\
& &{\hat \Sigma}_0={\hat {\cal S}}-\frac{1}{2}C_m\ .
\label{a53c}
\eeq
\esub
Then, we have 
\beq\label{a54}
{\hat \Sigma}_+\ket{C_m/2,s_0}=0\ . 
\eeq
The above indicates that the state $\ket{C_m/2,s_0}$ 
is the maximum weight state.

The commutation relations for ${\hat \Sigma}_{\pm,0}$ are given in the form 
\beq
& &[\ {\hat \Sigma}_0 \ , \ {\hat \Sigma}_{\pm}\ ]=\pm{\hat \Sigma}_{\pm}\ , 
\label{a55}\\
& &[\ {\hat \Sigma}_+ \ , \ {\hat \Sigma}_- \ ]=2\left({\hat \Sigma}_0+\Delta{\hat \Sigma}_0\right)\ , 
\label{a56}\\
%& &\Delta{\hat \Sigma}_0=\Delta{\hat \Sigma}_0^{(+)}-\Delta{\hat \Sigma}_0^{(-)}\ , \nonumber\\
& &\Delta{\hat \Sigma}_0=\frac{\epsilon}{2}\frac{\left({\hat {\cal S}}_0+{\hat {\cal S}}+1\right)\left({\hat {\cal S}}_0+C_m-{\hat {\cal S}}\right)}
{\frac{1}{2}C_m-{\hat {\cal S}}+\epsilon}\ . 
%, \nonumber\\
%& &\Delta{\hat \Sigma}_0^{(-)}=\frac{\epsilon}{2}\frac{\left({\hat {\cal S}}_0+{\hat {\cal S}}\right)\left({\hat {\cal S}}_0+C_m-{\hat {\cal S}}+1\right)}
%{{\hat {\cal S}}-s_{\rm min}+\epsilon}\ . 
\label{a57} 
\eeq
The operator ${\hat {\mib {\Sigma}}}^2$ corresponding to the Casimir operator of the $su(2)$-algebra is given as
\beq
{\hat {\mib \Sigma}}^2&=&
{\hat \Sigma}_0^2+\frac{1}{2}\left({\hat \Sigma}_-{\hat \Sigma}_++{\hat \Sigma}_+{\hat \Sigma}_-\right)
\nonumber\\
&=&{\hat {\Sigma}}\left({\hat \Sigma}+1\right)
-\left(\Delta{\hat \Sigma}_0^{(+)}+\Delta{\hat \Sigma}_0^{(-)}\right)\ , 
\label{a58}\\
{\hat \Sigma}&=&
{\hat {\cal S}}_0+\frac{1}{2}C_m\ . 
\label{a59}
\eeq
We can see that the set ${\hat \Sigma}_{\pm,0}$ forms a kind of the pseudo $su(2)$-algebra.

Thus, we could learn the existence of the raising and the lowering operator for the magnitude of the $su(2)$-spin. 
With the use of the operator ${\hat \Sigma}_+$ and ${\hat {\cal S}}_{\pm}$, we can express the state $\ket{s,s_0}$ in the form 
\bsub\label{3-13}
\beq
& &\ket{s,s_0}=\left({\hat {\cal S}}_+\right)^{s+s_0}\ket{s,-s}\ , 
\label{3-13a}\\
& &\ket{s,-s}=\left({\hat {\cal S}}_-{\hat \Sigma}_+\right)^{s-\rho}\ket{\rho,-\rho}\ , 
\label{3-13b}\\
& &\ket{\rho,-\rho}=({\hat b}^*)^{C_m-2\rho}\ket{0}\ . \qquad \left(\rho=0\ {\rm and}\ \frac{1}{2}\right)
\label{3-13c}
\eeq
\esub
In the above relations, we omitted the normalization constants. 
If $\rho=0$ and $1/2$, $s$ become integer and half-integer, respectively. 
However, we must remark that the above idea is not proper to our representation. 
In the case of the Schwinger representation, we obtain the state $\dket{s,s_0}$ by replacing ${\hat {\cal S}}_{\pm}$, ${\hat \Sigma}_+$ 
and $\ket{\rho,-\rho}$ with ${\hat S}_{\pm}$, ${\hat T}_+$ and $\dket{\rho,-\rho}=({\hat b}^*)^{\rho}\ket{0}$.

\setcounter{equation}{0}
\section{Connection to the Holstein-Primakoff representation}

It may be interesting to show how the Holstein-Primakoff representation \cite{3} 
can be derived from our representation (I.3.9). 
The three $su(2)$-generators (I.3.9) are rewritten to the following forms: 
\beq\label{4-1}
{\hat {\cal S}}_+={\hat a}^*{\hat \beta}^*\cdot\sqrt{2{\hat {\cal S}}-{\hat a}^*{\hat a}}\ , \qquad
{\hat {\cal S}}_-=\sqrt{2{\hat {\cal S}}-{\hat a}^*{\hat a}}\cdot{\hat \beta}{\hat a}\ , \qquad
{\hat {\cal S}}_0={\hat a}^*{\hat a}-{\hat {\cal S}}\ . 
\eeq
Here, (${\hat \beta}, {\hat \beta}^*)$ is defined in the form 
\beq\label{4-2}
{\hat \beta}=\left(\sqrt{{\hat b}^*{\hat b}+1+\epsilon}\right)^{-1}\cdot{\hat b}\ , \qquad
{\hat \beta}^*={\hat b}^*\cdot\left(\sqrt{{\hat b}^*{\hat b}+1+\epsilon}\right)^{-1}\ . 
\eeq
Further, for the above rewriting, we used the relation 
\beq\label{4-3}
C_m-{\hat b}^*{\hat b}=2{\hat {\cal S}}-{\hat a}^*{\hat a}\ . 
\eeq
The relation (\ref{4-3}) comes from the form (I.3.10). 
Any of $({\hat \beta}, {\hat \beta}^*)$ connects with any of $({\hat a}, {\hat a}^*)$ and we have the relation 
\bsub\label{4-4}
\beq
& &{\hat \beta}{\hat \beta}^*=1-\frac{\epsilon}{{\hat b}^*{\hat b}+1+\epsilon}\rightarrow 1\ , \quad (\epsilon\rightarrow 0)
\label{4-4a}\\
& &{\hat \beta}^*{\hat \beta}=1-\frac{\epsilon}{{\hat b}^*{\hat b}+\epsilon}\ . 
\label{4-4b}
\eeq
\esub
Therefore, $[{\hat \beta}\ , \ {\hat \beta}^*]$ is given as 
\beq\label{4-5}
[\ {\hat \beta}\ , \ {\hat \beta}^*\ ]=\frac{\epsilon}{{\hat b}^*{\hat b}+\epsilon}\ . 
\eeq
The operator $[{\hat \beta}\ , \ {\hat \beta}^*]$ plays a role of projection operator and, for the state (\ref{a50}), we have 
\beq\label{4-6}
[\ {\hat \beta}\ , \ {\hat \beta}^*\ ]\ket{s,s_0}=\frac{\epsilon}{C_m-s+s_0+\epsilon}\ket{s,s_0}=\delta_{s-s_0,C_m}\ket{s,s_0}\ . \qquad (\epsilon\rightarrow 0)
\eeq
Our boson representation is given in the space $P$ in Fig.\ref{fig:subspace} or Fig.\ref{fig:ss0} and, then, $\delta_{s-s_0,C_m}$ takes the value 1 in the case 
\bsub\label{4-7}
\beq
s=s_{\rm max}=\frac{C_m}{2}\ , \qquad
s_0=-s_{\rm max}=-\frac{C_m}{2}\ , \qquad
\delta_{s-s_0,C_m}\ket{s,s_0}=\ket{\frac{C_m}{2},-\frac{C_m}{2}}\ . 
\label{4-7a}
\eeq
In any other case, we have 
\beq
%\delta_{s-s_0,C_m}=0\ , \qquad 
\delta_{s-s_0,C_m}\ket{s,s_0}=0\ . 
\label{4-7b}
\eeq
\esub

With the use of $({\hat \beta},{\hat \beta}^*)$, we define the operator $({\hat c},{\hat c}^*)$: 
\beq\label{4-8}
{\hat c}={\hat \beta}{\hat a}\ (={\hat a}{\hat \beta})\ , \qquad
{\hat c}^*={\hat a}^*{\hat \beta}^*\ (={\hat \beta}^*{\hat a}^*)\ . 
\eeq
The operator $({\hat c},{\hat c}^*)$ satisfies the relation
\bsub\label{4-9}
\beq
{\hat c}{\hat c}^*&=&{\hat a}{\hat \beta}{\hat \beta}^*{\hat a}^*={\hat a}{\hat a}^*\ , 
\label{4-9a}\\
{\hat c}^*{\hat c}&=&{\hat a}^*{\hat \beta}^*{\hat \beta}{\hat a}={\hat a}^*{\hat a}\cdot({\hat \beta}^*{\hat \beta})={\hat a}^*{\hat a}
-{\hat a}^*{\hat a}[\ {\hat\beta}\ , \ {\hat \beta}^*\ ] \nonumber\\
&=&{\hat a}^*{\hat a}-\left({\hat {\cal S}}+{\hat {\cal S}}_0\right)[\ {\hat \beta}\ , \ {\hat \beta}^*\ ] \rightarrow {\hat a}^*{\hat a}\ . 
\qquad (\epsilon\rightarrow 0)
\label{4-9b}
\eeq
\esub
We notice that $({\hat {\cal S}}+{\hat {\cal S}}_0)[{\hat \beta} , {\hat \beta}^*]\ket{C_m/2,-C_m/2}=(C_m/2-C_m/2)\ket{C_m/2,-C_m/2}$ and 
for any other case, $[{\hat \beta} , {\hat \beta}^*]\ket{s,s_0}=0$. 
Therefore, $({\hat c}, {\hat c}^*)$ is boson operator:
\beq\label{4-10}
[\ {\hat c}\ , \ {\hat c}^*\ ]=[\ {\hat a}\ , \ {\hat a}^*\ ]=1\ , \qquad
{\hat c}\ket{0}={\hat b}\ket{0}={\hat a}\ket{0}=0\ . 
\eeq
Further, we obtain the relation 
\beq\label{4-11}
[\ {\hat c}\ , \ {\hat {\cal S}}\ ]=[\ {\hat c}^*\ , \ {\hat {\cal S}}\ ]=0\ . 
\eeq
With the use of the relations obtained above, 
${\hat {\cal S}}_{\pm,0}$ can be rewritten to the form 
\beq\label{4-12}
{\hat {\cal S}}_+={\hat c}^*\sqrt{2{\hat {\cal S}}-{\hat c}^*{\hat c}}\ , \qquad
{\hat {\cal S}}_-=\sqrt{2{\hat {\cal S}}-{\hat c}^*{\hat c}}\ {\hat c}\ , \qquad
{\hat {\cal S}}_0={\hat c}^*{\hat c}-{\hat {\cal S}}\ . 
\eeq

By setting $s+s_0=n$ in the state (\ref{a50}), we have 
\beq\label{4-13}
f\left({\hat {\cal S}}, {\hat c}^*{\hat c}\right)\ket{s,s_0}=f(s,n)\ket{s,s_0}=f(s,{\hat c}^*{\hat c})\ket{s,s_0}\ . 
\eeq
Therefore, in the space spanned by $\ket{s,s_0}$ $(s_0=-s,\ -s+1,\cdots ,\ s-1,\ s)$, 
the expression (\ref{4-12}) can be regarded as 
\beq\label{4-14}
{\hat S}_+={\hat c}^*\sqrt{2s-{\hat c}^*{\hat c}}_ , \qquad
{\hat S}_-=\sqrt{2s-{\hat c}^*{\hat c}}\ {\hat c}\ , \qquad
{\hat S}_0={\hat c}^*{\hat c}-s\ . 
\eeq
The above is nothing but the Holstein-Primakoff representation.
In Ref.\citen{4}, we discussed an idea how to derive the Holstein-Primakoff representation from the Schwinger one. 
In this case, much more lengthy discussion was necessary. 
The main reason may be attributed to the fact that the operator corresponding to 
$({\hat c}, {\hat c}^*)$ does not satisfy the simple boson commutation relation. 
The state (\ref{a50}) can be rewritten to 
\beq
& &\ket{s,s_0}~\frac{1}{\sqrt{(s+s_0)!}}({\hat c}^*)^{s+s_0}\ket{s,-s}\ ,
\label{4-15}\\
& &\ket{s,-s}=\frac{1}{\sqrt{(C_m-2s)!}}({\hat b}^*)^{C_m-2s}\ket{0}\ (=\ket{\phi}) \ . 
\label{4-16}
\eeq
Clearly, ${\hat c}$ satisfies 
\beq\label{4-17}
{\hat c}\ket{\phi}=0\ . 
\eeq
Since the state $\ket{\phi}$ is the vacuum of ${\hat c}$, we have the orthogonal set: 
\beq\label{4-18}
\ket{n}=\frac{1}{\sqrt{n!}}({\hat c}^*)^n\ket{\phi}\ . \qquad (n=0,\ 1,\cdots ,\ 2s)
\eeq
The above is the connection to the Holstein-Primakoff representation.

\setcounter{equation}{0}

\section{Connection to the Schwinger representation}

In this section, we will discuss how the Schwinger boson representation is connected to the present one. 
The three generators ${\hat {\cal S}}_{\pm,0}$ and ${\hat {\cal S}}$ given 
in the relations (I.3.9) and (I.3.10) are 
rewritten to the form 
\beq
& &{\hat {\cal S}}_+={\hat a}^*{\hat \beta}^*\cdot\sqrt{C_m-{\hat b}^*{\hat b}}\ , \quad
{\hat {\cal S}}_-=\sqrt{C_m-{\hat b}^*{\hat b}}\cdot{\hat \beta}{\hat a}\ , \quad
{\hat {\cal S}}_0=\frac{1}{2}({\hat a}^*{\hat a}-(C_m-{\hat b}^*{\hat b}))\ , \qquad
\label{5-1}\\
& &{\hat {\cal S}}=\frac{1}{2}({\hat a}^*{\hat a}+(C_m-{\hat b}^*{\hat b}))\ . 
\label{5-2}
\eeq
Here, $({\hat \beta}, {\hat \beta}^*)$ is given in the relation (\ref{4-2}). 
In order to rewrite the expressions (\ref{5-1}) and (\ref{5-2}), we introduce the operator $({\maru b}, {\maru b}^*)$ in the form 
\beq\label{5-3}
{\maru b}={\hat \beta}^*\sqrt{C_m-{\hat b}^*{\hat b}}\ , \qquad
{\maru b}^*=\sqrt{C_m-{\hat b}^*{\hat b}}\ {\hat \beta}\ .
\eeq
With the use of $({\maru b}, {\maru b}^*)$, ${\hat {\cal S}}_{\pm,0}$ and ${\hat {\cal S}}$ can be expressed as 
\beq
& &{\maru {\cal S}}_+={\hat a}^*{\maru b}\ , \qquad
{\maru {\cal S}}_-={\maru b}^*{\hat a}\ , \qquad
{\maru {\cal S}}_0=\frac{1}{2}({\hat a}^*{\hat a}-{\maru b}^*{\maru b})\ , 
\label{5-4}\\
& &{\maru {\cal S}}=\frac{1}{2}({\hat a}^*{\hat a}+{\maru b}^*{\maru b})\ . 
\label{5-5}
\eeq
Here, it should be noted that we rewrite ${\hat {\cal S}}_{\pm,0}$ and ${\hat {\cal S}}$ shown in the relations 
(\ref{5-1}) and (\ref{5-2}), in the 
forms (\ref{5-4}) and (\ref{5-5}) and, then, ${\maru {\cal S}}_{\pm,0}={\hat {\cal S}}_{\pm,0}$ and ${\maru {\cal S}}={\hat {\cal S}}$. 
If $({\maru b}, {\maru b}^*)$ is boson operator, the expression (\ref{5-4}) (and (\ref{5-5})) reduce to the 
Schwinger boson representation (I.3.1) (and (I.3.2)). 
Therefore, it may be interesting to investigate the condition, under which $({\maru b}, {\maru b}^*)$ can be 
regarded as boson. 
For this aim, we must calculate the commutation relation $[{\maru b},{\maru b}^*]$, the result of which 
is as follows: 
\beq\label{5-6}
[\ {\maru b}\ , \ {\maru b}^*\ ]&=&
1-(C_m+1-{\hat b}^*{\hat b})[\ {\hat \beta}\ , \ {\hat \beta}^*\ ]
=1-(C_m+1-{\hat b}^*{\hat b})\frac{\epsilon}{{\hat b}^*{\hat b}+\epsilon} 
\nonumber\\
&=&1-\frac{\epsilon({\maru b}^*{\maru b}+1)}{C_m-{\maru b}^*{\maru b}+\epsilon}\ . 
\eeq
Here, we used the relation 
\beq\label{5-7}
{\maru b}{\maru b}^*=(C_m+1-{\maru b}^*{\maru b}){\hat \beta}^*{\hat \beta}\ , \qquad
{\maru b}^*{\maru b}=C_m-{\hat b}^*{\hat b}\ . 
\eeq
In the relation (\ref{5-6}), we can see that if $C_m\rightarrow \infty$, 
$({\maru b},{\maru b}^*)$ may be regarded as boson.

In order to clarify the above-mentioned situation, we consider orthogonal set constructed by ${\maru b}^*$. 
First, we introduce the state $\kket{0}$ defined as 
\beq\label{5-8}
\kket{0}=({\hat \beta}^*)^{C_m}\ket{0}=\frac{1}{\sqrt{C_m!}}({\hat b}^*)^{C_m}\ket{0}\ . 
\eeq
The state $\kket{0}$ is the vacuum for ${\maru b}$:
\beq\label{5-9}
{\maru b}\kket{0}={\hat \beta}^*\sqrt{C_m-{\hat b}^*{\hat b}}\cdot\frac{1}{\sqrt{C_m!}}({\hat b}^*)^{C_m}\ket{0}=0\ . 
\eeq
Then, we define the state $\kket{n}$ in the form 
\beq\label{5-10}
\kket{n}=({\hat \beta}^*)^{C_m-n}\ket{0}=\frac{1}{\sqrt{(C_m-n)!}}({\hat b}^*)^{C_m-n}\ket{0}\ . 
\qquad
(n=0,\ 1,\ 2,\cdots ,\ C_m)
\eeq
It can be proved that $\kket{n}$ is expressed as 
\beq\label{5-11}
\kket{n}=\frac{1}{\sqrt{n!}}({\maru b}^*)^n\kket{0}\ . 
\eeq
If $n=C_m$, we have the following relation:
\beq\label{5-12}
{\maru b}^*\kket{C_m}=\sqrt{C_m-{\hat b}^*{\hat b}}\ {\hat \beta}\ket{0}=0\ . 
\eeq
It may be interesting to see that the operation of ${\maru b}^*$ on $\kket{C_m}$ vanishes. 
Therefore, if $C_m\rightarrow \infty$, the relation (\ref{5-12}) becomes meaningless. 
The commutation relation (\ref{5-6}) gives us the following: 
\bsub\label{5-13}
\beq
& &[\ {\maru b}\ , \ {\maru b}^*\ ]\kket{n}=\kket{n}\ , \qquad (n=0,\ 1,\ 2,\cdots , \ C_m-1)
\label{5-13a}\\
& &[\ {\maru b}\ , \ {\maru b}^*\ ]\kket{C_m}=-C_m\kket{C_m}\ .
\label{5-13b}
\eeq
\esub
In the relation (\ref{5-13b}), we have ${\maru b}{\maru b}^*\kket{C_m}=0$ and 
${\maru b}^*{\maru b}\kket{C_m}=C_m\kket{C_m}$, which comes from the relation (\ref{5-12}). 
From the above argument, we can conclude that if $C_m\rightarrow \infty$, $({\maru b},{\maru b}^*)$ can be regarded as 
boson operator. 
In this connection, we mention that if $C_m=1$, $({\maru b},{\maru b}^*)$ becomes fermion operator.

Under the above consideration, we investigate the eigenvalue problem for ${\maru {\cal S}}_{\pm,0}$ and ${\maru {\cal S}}$. 
First, we introduce the state $\kket{s}$ in the form 
\beq\label{5-14}
\kket{s}=\frac{1}{\sqrt{(2s)!}}({\maru b}^*)^{2s}\kket{0}\ . 
\eeq
Here, it should be noted that $s$ takes the values 
\beq\label{5-15}
s=0,\ 1/2,\ 1,\ 3/2,\cdots ,\ (C_m-1)/2,\ C_m/2\ . 
\eeq
In the Schwinger representation, $s=0,\ 1/2,\cdots ,\ \infty$. 
This difference has been mentioned in \S 1. 
The state $\kket{s}$ is the minimum weight state satisfying the relation 
\beq\label{5-16}
{\maru {\cal S}}_-\kket{s}=0\ , \qquad
{\maru {\cal S}}_0\kket{s}=-s\kket{s}\ , \qquad
{\maru {\cal S}}\kket{s}=s\kket{s}\ . 
\eeq
Then, we define the following state:
\beq\label{5-17}
\kket{s,s_0}=\sqrt{\frac{(s-s_0)!}{(2s)!(s+s_0)!}}\left({\maru {\cal S}}_+\right)^{s+s_0}\dket{s}\ . 
\eeq
Together with the properties, $\kket{s,s_0}$ can be shown in the form 
\beq
& &\kket{s,s_0}=\frac{1}{\sqrt{(s+s_0)!(s-s_0)!}}({\hat a}^*)^{s+s_0}({\maru b}^*)^{s-s_0}\kket{0}\ , 
\label{5-18}\\
& &{\maru {\cal S}}\kket{s,s_0}=s\kket{s,s_0}\ , \qquad
{\maru {\cal S}}_0\kket{s,s_0}=s_0\kket{s,s_0}\ . 
\label{5-19}
\eeq
Of course, $s=0,\ 1/2,\ 1,\ 3/2,\cdots , \ (C_m-1)/2,\ C_m/2$ and for a given $s$, 
$s_0=-s,\ -s+1, \cdots,\ s-1,\ s$.

The commutation relations among ${\maru {\cal S}}_{\pm,0}$ and ${\maru {\cal S}}$ are enumerated as follows: 
\bsub\label{5-20}
\beq
& &[\ {\maru {\cal S}}_+\ , \ {\maru {\cal S}}_-\ ]=2{\maru {\cal S}}_0-{\hat a}^*{\hat a}(1-[\ {\maru b}\ , \ {\maru b}^*\ ])\ , 
\label{5-20a}\\
& &[\ {\maru {\cal S}}_0\ , \ {\maru {\cal S}}_-\ ]=-{\maru {\cal S}}_-+\frac{1}{2}{\maru b}^*{\hat a}(1-[\ {\maru b}\ , \ {\maru b}^*\ ])\ , 
\label{5-20b}\\
& &[\ {\maru {\cal S}}\ , \ {\maru {\cal S}}_-\ ]=-\frac{1}{2}{\maru b}^*{\hat a}(1-[\ {\maru b}\ , \ {\maru b}^*\ ])\ .
\label{5-20c}
\eeq
Of course, we have 
\beq
& &[\ {\maru {\cal S}}_0\ , \ {\maru {\cal S}}_+\ ]={\maru {\cal S}}_+-\frac{1}{2}(1-[\ {\maru b}\ , \ {\maru b}^*\ ]){\hat a}^*{\maru b}\ , 
\label{5-20d}\\
& &[\ {\maru {\cal S}}\ , \ {\maru {\cal S}}_+\ ]=\frac{1}{2}(1-[\ {\maru b}\ , \ {\maru b}^*\ ]){\hat a}^*{\maru b}\ . 
\label{5-20e}
\eeq
\esub
With the use of the relations (\ref{5-10}) and (\ref{5-13}), we have 
\beq\label{5-21}
{\hat a}(1-[\ {\maru b}\ , \ {\maru b}^*\ ])\kket{n}=0\ . \qquad (n=0,\ 1,\ 2,\cdots ,\ C_m-1,\ C_m)
\eeq
Therefore, we can conclude that ${\maru {\cal S}}_{\pm,0}$ obey the $su(2)$-algebra and commute with ${\maru {\cal S}}$.

In last discussion, the connection of the Schwinger representation to ours was clarified. 
We continue this discussion by introducing a new boson space 
composed of boson $({\ma b},{\ma b}{}^*)$. 
Of course, the orthogonal set is given by 
\beq\label{5-22}
\rrket{n}=\frac{1}{\sqrt{n!}}({\ma b}{}^*)^n\rrket{0}\ . \qquad
(n=0,\ 1,\ 2,\cdots)
\eeq
In this space, the following operator is introduced: 
\beq\label{5-23}
& &({\maru b})_c=\left(\sqrt{C_m-{\ma b}{}^*{\ma b}+\epsilon}\right)^{-1}\cdot\sqrt{C_m-{\ma b}{}^*{\ma b}}\cdot {\ma b}\ , \nonumber\\
& &({\maru b}^*)_c={\ma b}{}^*\cdot\sqrt{C_m-{\ma b}{}^*{\ma b}}\cdot\left(\sqrt{C_m-{\ma b}{}^*{\ma b}+\epsilon}\right)^{-1}\ . 
\eeq
We can easily verified the relation 
\beq\label{5-24}
& &({\maru b})_c\rrket{0}=0\ , 
\label{5-24}\\
& &({\maru b}^*)_c\rrket{n}={\ma b}{}^*\rrket{n}\ , \quad (n=0,\ 1,\ 2,\cdots ,\ C_m-1)\ , \qquad
({\maru b}^*)_c\rrket{C_m}=0\ . 
\eeq
Further, we have 
\beq\label{5-26}
[\ ({\maru b})_c\ , \ ({\maru b}^*)_c\ ]=1-
\frac{({\ma b}{}^*{\ma b}+1)\epsilon}{C_m-{\ma b}{}^*{\ma b}+\epsilon}+\frac{{\ma b}{}^*{\ma b}\epsilon}{C_m-{\ma b}{}^*{\ma b}+1+\epsilon}\ . 
\eeq
This relation leads us to 
\bsub\label{5-27}
\beq
& &[\ ({\maru b})_c\ , \ ({\maru b}^*)_c\ ]\rrket{n}=\rrket{n}\ , \qquad (n=0,\ 1,\ 2,\cdots ,\ C_m-1)
\label{5-27a}\\
& &[\ ({\maru b})_c\ , \ ({\maru b}^*)_c\ ]\rrket{C_m}=-C_m\rrket{C_m}\ .
\label{5-27b}
\eeq
\esub
All of the above relations suggest us that $(({\maru b})_c, ({\maru b}^*)_c)$ may be regarded as the counterpart of $({\maru b},{\maru b}^*)$ 
in the boson space composed of $({\ma b},{\ma b}{}^*)$.

Next, we consider the counterparts of ${\maru {\cal S}}_{\pm,0}$ and ${\maru {\cal S}}$ in the space 
composed of $({\hat a},{\hat a}^*)$ and $({\ma b},{\ma b}{}^*)$, which are denoted as 
$({\maru {\cal S}}_{\pm,0})_c$ and $({\maru {\cal S}})_c$. 
In this space, we introduce the following set: 
\beq\label{5-28}
& &\rrket{s,s_0}=\frac{1}{\sqrt{(s+s_0)!(s-s_0)!}}({\hat a}^*)^{s+s_0}({\ma b}{}^*)^{s-s_0}\rrket{0}\ . 
\quad (s=0,\ 1/2,\ 1,\ 3/2, \cdots,\ C_m/2)\nonumber\\
& &
\eeq
Further, $({\maru {\cal S}}_{\pm,0})_c$ and $({\maru {\cal S}})_c$ are introduced in the form 
\bsub\label{5-29}
\beq
& &({\maru {\cal S}}_+)_c={\hat a}^*({\maru b})_c=\left(\sqrt{C_m-{\ma b}{}^*{\ma b}+\epsilon}\right)^{-1}\cdot\sqrt{C_m-{\ma b}{}^*{\ma b}}\cdot{\hat a}^*{\ma b}\ , 
\label{5-29a}\\
& &({\maru {\cal S}}_-)_c=({\maru b}^*)_c{\hat a}={\ma b}{}^*{\hat a}\cdot\sqrt{C_m-{\ma b}{}^*{\ma b}}\cdot\left(\sqrt{C_m-{\ma b}{}^*{\ma b}+\epsilon}\right)^{-1}\ , 
\label{5-29b}\\
& &({\maru {\cal S}}_0)_c=\frac{1}{2}({\hat a}^*{\hat a}-({\maru b}^*)_c({\maru b})_c)
=\frac{1}{2}({\hat a}^*{\hat a}-{\ma b}{}^*{\ma b})+\frac{1}{2}\Delta {\ma {\cal S}}\ , 
\label{5-29c}
\eeq
\esub
\beq\label{5-30}
({\maru {\cal S}})_c=\frac{1}{2}({\hat a}^*{\hat a}+({\maru b}^*)_c({\maru b})_c)
=\frac{1}{2}({\hat a}^*{\hat a}+{\ma b}{}^*{\ma b})-\frac{1}{2}\Delta {\ma {\cal S}}\ . 
\eeq
Here, $\Delta {\ma {\cal S}}$ is given as 
\beq\label{5-31}
\Delta{\ma {\cal S}}=\frac{\epsilon {\ma b}{}^*{\ma b}}{C_m-{\ma b}{}^*{\ma b}+1+\epsilon}\ . 
\eeq
Since, in the present space, $\Delta {\ma {\cal S}}$ is positive-definite, it can be omitted and, then, $\rrket{s,s_0}$ is the eigenstate of 
$({\maru {\cal S}})_c$ and $({\maru {\cal S}}_0)_c$ with the eigenvalues $s$ and $s_0$, respectively. 
Therefore, the following commutation relations are easily verified: 
\beq\label{5-32}
[\ ({\maru {\cal S}}_0)_c\ , \ ({\maru {\cal S}}_{\pm})_c\ ]=\pm({\maru {\cal S}}_{\pm})_c\ , \qquad
[\ ({\maru {\cal S}})_c\ , \ ({\maru {\cal S}}_{\pm,0})_c\ ]=0\ . 
\eeq
The commutation relation $[({\maru {\cal S}}_+)_c , ({\maru {\cal S}}_-)_c ]$ is given by 
\beq
& &[\ ({\maru {\cal S}}_+)_c\ , \ ({\maru {\cal S}}_-)_c\ ]=2\left( ({\maru {\cal S}}_{0})_c-\Delta({\maru {\cal S}}_0)_c\right)\ , 
\label{5-33}\\
& &\Delta({\maru {\cal S}}_0)_c=\frac{1}{2}\left(
\frac{\epsilon{\hat a}^*{\hat a}({\ma b}{}^*{\ma b}+1)}{C_m-{\ma b}{}^*{\ma b}+\epsilon}-\frac{\epsilon({\hat a}^*{\hat a}+1){\ma b}{}^*{\ma b}}{C_m-
{\ma b}{}^*{\ma b}+1+\epsilon}\right)\ . 
\label{5-34}
\eeq
We can show that the operation of $\Delta ({\maru {\cal S}}_0)_c$ on the state $\rrket{s,s_0}$ 
makes the result of the operation vanish. 
Further, we can prove the relation 
\beq\label{5-35}
({\maru {\mib {\cal S}}}^2)_c&=&
({\maru {\cal S}}_0)_c^2+\frac{1}{2}\left(({\maru {\cal S}}_-)_c({\maru {\cal S}}_+)_c+({\maru {\cal S}}_+)_c({\maru {\cal S}}_-)_c\right)\nonumber\\
&=&({\hat {\cal S}})_c\left(({\hat {\cal S}})_c+1\right)\ . 
\eeq
Thus, we can conclude that $({\maru {\cal S}}_{\pm,0})_c$ obey the $su(2)$-algebra 
with the magnitude of the $su(2)$-spin $({\maru {\cal S}})_c$.

\setcounter{equation}{0}
\section{Concluding remarks}

In this paper, we investigated various theoretical features of the 
new boson representation of the $su(2)$-algebra presented in Part I. 
As was stressed in Part I, essential difference between the Schwinger representation and ours can be 
found in the expressions of the operators which give the quantum numbers specifying the orthogonal set, $(s, s_0)$. 
They are completely opposite to each other. 
Therefore, we must put each representation to its proper use. 
As concluding remarks, we will mention two points.

First point is concerned with the promise mentioned at the end of \S 2. 
We will examine the subspaces $R_p$, $R_q$ and $R$. 
The case $R$ is simple: 
the $su(1,1)$-algebra presented in \S 2 for the case $t\geq \mu+1/2$. 
The cases $R_p$ and $R_q$ are a little bit complicated. 
As a preliminary argument, we consider the case of the Holstein-Primakoff representation (4.14).  
Let us investigate the case in which the order of $2s$ and ${\hat c}^*{\hat c}$ in ${\hat S}_{\pm}$ is changed. 
In this case, we define the operators ${\maru {\rm T}}_{\pm,0}$: 
\beq\label{a40}
{\maru {\rm T}}_+={\hat c}^*\cdot\sqrt{{\hat c}^*{\hat c}-2s}\ , \quad
{\maru {\rm T}}_-=\sqrt{{\hat c}^*{\hat c}-2s}\cdot{\hat c}\ , \quad
{\maru {\rm T}}_0={\hat c}^*{\hat c}-s\ \left(={\hat S}_0\right)\ . 
\eeq
The commutation relations are given in the form 
\beq\label{a41}
[\ {\maru {\rm T}}_+\ , \ {\maru {\rm T}}_-\ ]=-2{\maru {\rm T}}_0\ , \qquad
[\ {\maru {\rm T}}_0\ , \ {\maru {\rm T}}_{\pm}\ ]=\pm{\maru {\rm T}}_{\pm}\ . 
\eeq
Of course, we have the state 
\beq\label{a42}
& &\ket{n}=\left({\maru {\rm T}}_+\right)^{n-(2s+1)}\ket{2s+1}=({\hat c}^*)^n\ket{0}\ , 
\nonumber\\
& &{\maru {\rm T}}_-\ket{2s+1}=0\ , \qquad n=2s+1,\ 2s+2, \cdots \ .
\eeq
The Casimir operator is given as 
\beq\label{a43}
{\maru {\mib {\rm T}}}^2=(s+1)((s+1)-1)\ . 
\eeq
Here, the quantity $(s+1)$ indicates the magnitude of the $su(1,1)$-spin. 
In this case, we obtain the $su(1,1)$-algebra.

Following the above idea, the order of ${\hat T}_m$ and ${\hat T}_0$ in the relations (I.3.7b) is changed. 
In this case, we define the operators ${\maru {\cal T}}_{\pm,0}$: 
\beq\label{a44}
& &{\maru {\cal T}}_+={\hat T}_+\cdot\sqrt{{\hat T}_0-{\hat T}_m}\cdot\left(\sqrt{{\hat T}_0+{\hat T}+\epsilon}\right)^{-1}\ , \nonumber\\
& &{\maru {\cal T}}_-=\left(\sqrt{{\hat T}_0+{\hat T}+\epsilon}\right)^{-1}\cdot\sqrt{{\hat T}_0-{\hat T}_m}\cdot{\hat T}_-\  , \nonumber\\
& &{\maru {\cal T}}_0={\hat T}_0-\frac{1}{2}\left({\hat T}_m+{\hat T}\right)\ \left(={\hat {\cal S}}_0\right)\ . 
\eeq
In the case ${\hat T}_m=C_m+1-{\hat T}$, ${\hat T}_0-{\hat T}_m={\hat T}_0+{\hat T}-(C_m+1)$. 
The commutation relation $[\ {\maru {\cal T}}_+\ , \ {\maru {\cal T}}_-\ ]$ is given in the form 
\beq\label{a45}
& &[\ {\maru {\cal T}}_+\ , \ {\maru {\cal T}}_-\ ]=-2\left({\maru {\cal T}}_0+\Delta{\maru {\cal T}}_0\right)\ , 
\nonumber\\
& &\Delta{\maru {\cal T}}_0=\Delta{\maru {\cal T}}_0^{(+)}-\Delta{\maru {\cal T}}_0^{(-)}\ , \qquad
\Delta{\maru {\cal T}}_0^{(+)}=\Delta{\hat {\cal S}}_0^{(+)}\ , \qquad
\Delta{\maru {\cal T}}_0^{(-)}=\Delta{\hat {\cal S}}_0^{(-)}\ . 
\eeq
Here, $\Delta{\hat {\cal S}}_0^{(\pm)}$ is given in the relation (\ref{a25}). 
Since in the spaces $R_p$ and $R_q$, we have $t_0+t>0$ and $t_0+t-1>0$, 
that is, ${\hat T}_0+{\hat T}$ and ${\hat T}_0+{\hat T}-1$ are positive-definite and, then, 
$\Delta{\maru {\cal T}}_0\rightarrow 0$, $(\epsilon\rightarrow 0)$. 
Therefore, together with the other, we have 
\beq\label{a46}
[\ {\maru {\cal T}}_+\ , \ {\maru {\cal T}}_-\ ]=-2{\maru {\cal T}}_0\ , \qquad
[\ {\maru {\cal T}}_0\ , \ {\maru {\cal T}}_{\pm}\ ]=\pm{\maru {\cal T}}_{\pm}\ . 
\eeq
The above shows that in the spaces $R_p$ and $R_q$, ${\maru {\cal T}}_{\pm,0}$ forms 
the $su(1,1)$-algebra. 
In the same argument as that of $\Delta{\maru {\cal T}}_0$, the Casimir operator is expressed as 
\beq
& &{\maru {\mib {\cal T}}}^2={\maru {\cal T}}_0^2-\frac{1}{2}\left({\maru {\cal T}}_-{\maru {\cal T}}_+ + {\maru {\cal T}}_+{\maru {\cal T}}_-\right)
={\maru {\cal T}}\left({\maru {\cal T}}-1\right)\ , 
\label{a47}\\
& &{\maru {\cal T}}=\frac{1}{2}\left({\hat T}_m-{\hat T}\right)+1\ , \qquad
[\ {\maru {\cal T}}\ , \ {\maru {\cal T}}_{\pm,0}\ ]=0\ . 
\label{a48_1} 
\eeq
The orthogonal set should obey the condition $t_0 \geq t_m+1$ and the minimum weight state is given in the form 
\beq\label{a49_1}
R_p\ {\rm and}\ R_q\ ; \ \ \kket{t,t_m+1}\ , \quad {\maru {\cal T}}_-\kket{t,t_m+1}=0\ . 
\eeq
We can see that the above is completely in the same situation as that in the Holstein-Primakoff representation.

\begin{figure}[t]
\begin{center}
\includegraphics[height=9.5cm]{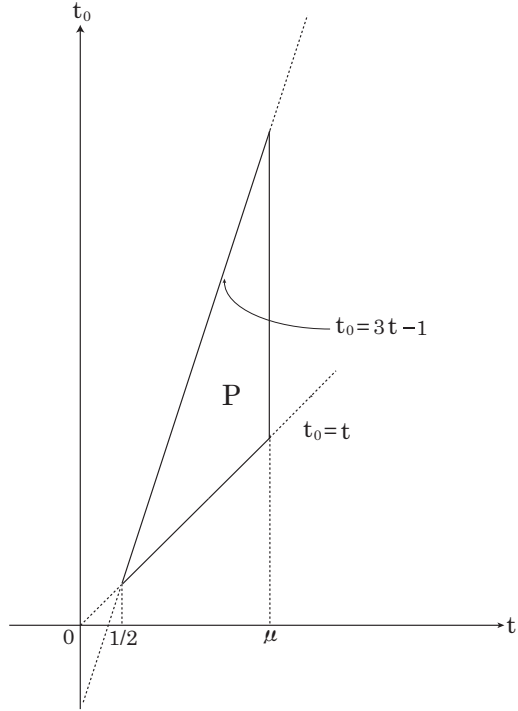}
\caption{The space $P$ is shown on the $t$-$t_0$ plane.
}
\label{fig:fig3}
\end{center}
\end{figure}

Second point is related to another example of ${\hat T}_m$. 
We consider the form 
\beq\label{6-11}
{\hat T}_m=3{\hat T}-1\ . 
\eeq
In this case, ${\hat {\cal S}}_{\pm,0}$ are expressed in the form 
\bsub\label{6-12}
\beq
{\hat {\cal S}}_+&=&{\hat T}_+\cdot\sqrt{3{\hat T}-1-{\hat T}_0}\cdot\left(\sqrt{{\hat T}_0+{\hat T}+\epsilon}\right)^{-1}\nonumber\\
&=&{\hat a}^*{\hat b}^*\cdot\sqrt{{\hat b}^*{\hat b}-2{\hat a}^*{\hat a}}\cdot\left(\sqrt{{\hat b}^*{\hat b}+1+\epsilon}\right)^{-1}\ , 
\label{6-12a}\\
{\hat {\cal S}}_-&=&\left(\sqrt{{\hat T}_0+{\hat T}+\epsilon}\right)^{-1}\cdot\sqrt{3{\hat T}-1-{\hat T}_0}\cdot{\hat T}_-\nonumber\\
&=&\left(\sqrt{{\hat b}^*{\hat b}+1+\epsilon}\right)^{-1}\cdot\sqrt{{\hat b}^*{\hat b}-2{\hat a}^*{\hat a}}\cdot{\hat b}{\hat a}\ , 
\label{6-12b}\\
{\hat {\cal S}}_0&=&{\hat T}_0-2{\hat T}+\frac{1}{2}=\frac{1}{2}(3{\hat a}^*{\hat a}-{\hat b}^*{\hat b})\ , 
\label{6-12c}
\eeq
\esub
\beq\label{6-13}
{\hat {\cal S}}&=&{\hat T}-\frac{1}{2}=\frac{1}{2}(-{\hat a}^*{\hat a}+{\hat b}^*{\hat b})\ . 
\qquad\qquad\qquad
\eeq
The expressions (\ref{6-12}) and (\ref{6-13}) obey the $su(2)$-algebra in the space $P$ shown 
in Fig.{\ref{fig:fig3}}. 
The quantum numbers $(s, s_0)$ are related to $(t,t_0)$ in the space $P$ as follows: 
\beq
& &s=t-\frac{1}{2}\ , \qquad s_0=t_0-2t+\frac{1}{2}\ , 
\label{6-14}\\
& &t=\frac{1}{2},\ 1,\ \frac{3}{2}, \cdots ,\ \mu-1,\ \mu\ . 
\label{6-15}
\eeq
The eigenstate of ${\hat {\cal S}}$ and ${\hat {\cal S}}_0$ is given as 
\beq\label{6-16}
\ket{s,s_0}&=&
\sqrt{\frac{(2s)!}{(s+s_0)!(3s+s_0)!}}\left({\hat T}_+\right)^{s+s_0}\cdot\frac{1}{\sqrt{(2s)!}}({\hat b}^*)^{2s}\ket{0}
\nonumber\\
&=&\frac{1}{\sqrt{(s+s_0)!(3s+s_0)!}}({\hat a}^*)^{s+s_0}({\hat b}^*)^{3s+s_0}\ket{0}\ . 
\eeq
Apparently, the forms (\ref{6-12}) and (\ref{6-13}) are very different from 
the forms (I.3.9) and (I.3.10). 
However, the form (\ref{6-12}) is rewritten to 
\beq\label{6-17}
{\hat {\cal S}}_+={\hat a}^*{\hat \beta}^*\cdot\sqrt{2{\hat {\cal S}}-{\hat a}^*{\hat a}}\ , \qquad
{\hat {\cal S}}_-=\sqrt{2{\hat {\cal S}}-{\hat a}^*{\hat a}}\cdot{\hat \beta}{\hat a}\ , 
\qquad
{\hat {\cal S}}_0={\hat a}^*{\hat a}-{\hat {\cal S}}\ . 
\eeq
The above leads us to the expression (4.12).

%%%%%%%%%%%%%%%%%%%%%%%%%%%%%%%%%%%%%%%%%%%%%%%%%%%%%%%%%%%%%%%%%%%%%%
%%%%%%%%%%%%%%%%%%%%%%%%%%%%%%%%%%%%%%%%%%%%%%%%%%%%%%%%%%%%%%%%%%%%%%%%

Through the above consideration, we may conjecture that there would exist various boson 
representations of the $su(2)$-algebra.

\section*{Acknowledgment}

One of the authors (M.Y.) would like to express his sincere thanks to Mrs. Y. Miyamoto 
for her cordial encouragement.
One of the authors (Y.T.) is partially supported by the Grants-in-Aid of the Scientific Research 
(No.23540311, No.26400277) from the Ministry of Education, Culture, Sports, Science and 
Technology in Japan. 

\appendix


%\bibliographystyle{ptephy}
%\bibliography{sample}
%
% once the .bbl file has been generated then place the text in your article.

%\vfill\pagebreak

\end{document}